\documentstyle[12pt]{article}
\begin{document}
\renewcommand{\baselinestretch}{1.5}
\newcommand{\ben}{\begin{equation}}
\newcommand{\een}{\end{equation}}
\newcommand{\nn}{\nonumber}
\newcommand{\bea}{\begin{eqnarray}}
\newcommand{\eea}{\end{eqnarray}}
\newcommand{\ra}{\rightarrow}
\newcommand{\eps}{\epsilon}
\newcommand{\epsp}{\epsilon'}

\begin{titlepage}
\begin{flushright}
IMSc/92-16

\today
\end{flushright}

\begin{center}
\Large {\bf CP-violating Phenomena in an \\
$ SU(2)_L \times SU(2)_R
\times U(1)_{B-L} \times SU(3)_H^{VL}$ \\
Horizontal Symmetric Model}

\vspace{0.4cm}

\normalsize
Debasis Bhowmick and Asim K. Ray

{\it Department of Physics, Visva-Bharati, Santiniketan, 731 235, India}

Sreerup Raychaudhuri

{\it Department of Pure Physics, University of Calcutta, \\
92 Acharya Prafulla Chandra Road, Calcutta 700 009, India}

and

S. Uma Sankar

{\it Institute of Mathematical Sciences,
Taramani, Madras 600 113, India.}
\end{center}
\vspace{1cm}

\begin{center}
{\bf Abstract}
\end{center}

We consider an $SU(2)_L \times SU(2)_R \times U(1)_{B-L} \times SU(3)_H^{VL}$
gauge model with natural flavour conservation in the Higgs sector, in which
CP-violation occurs due to the horizontal interactions only.  We
calculate the CP-violating observables $\epsilon$ and $\epsilon'$
of the neutral kaon
sector and $d_n$, the electric dipole moment of the neutron. The regions of the
parameter space which yield a value of $\epsilon$ that is in agreement with the
experiment, lead to predictions for $\epsilon'$ and $d_n$ which are at least
five
orders of magnitude smaller than the current experimental upper bounds.
\vspace{0.2cm}

\begin{flushleft}
(PACS Numbers: 11.30 Er, 11.30 Hv, 12.15 Ff, 12.15 Ji)
\end{flushleft}

\end{titlepage}

\begin{center} {\bf 1. Introduction} \end{center}

Ever since its discovery CP-violation \cite{ccft} has proved to be a very
intriguing and elusive phenomenon. Even after twenty five years of active
search it has been observed only in the neutral kaon sector so far. All
the three experimental observations, which show that CP is not conserved,
can be accounted for by a single parameter $\eps$, which is the ratio of
the imaginary to the real parts of the off-diagonal term in the $K^0 -
\bar{K}^0$ mass matrix \cite{cprev}.
The immensely successful standard model (SM) of
electroweak interactions attributes CP-violation to a complex phase $\delta$ in
the quark mixing (Kobayashi-Maskawa) matrix \cite{km} occuring in the charged
current interactions. This has two immediate consequences.  Firstly,
CP-violation cannot occur at tree level and can only occur due to diagrams with
one or more loops. Thus the SM accounts for the smallness of CP-violation
that is observed. Secondly, all the CP-violating observables can be calculated
in terms of $\delta$ and once $\delta$ has been determined by one of the
observables, one can predict all other CP-violating observables. In addition to
the three observed phenomena, the SM does predict other instances of
CP-violation, such as direct CP-violation in $K^0$ decays
(characterized by $\epsp$), CP-violation in the decays of
charged and neutral B-mesons and a very small neutron electric dipole moment
$d_n$.  Unfortunately, the currently obtainable experimental accuracy is
not enough to measure the SM predictions of these quantities.
Current experiments are searching to observe a non-zero value for
$\epsp$ \cite{epspexp}. CP-violating phenomena in B-mesons are
expected to be observable at the forthcoming $B$ factories \cite{bfac}.
It is hoped that these experiments would provide stringent tests
not only for the SM but
for the various non-standard scenarios whose predictions for
new CP-violating effects are
just below the current experimental accuracy.

An outstanding problem of particle physics, which is not explained by the SM,
is the fermion family replication. Related to this are the two other problems
of fermion mass hierarchy and the quark mixing angles. Horizontal symmetries
between fermion generations have been postulated to account for the
multiplicity of fermion doublets observed in nature. When these symmetries are
made local, the resulting gauge bosons couple to fermions of different
generations which leads to flavour changing neutral currents at tree level.
Therefore the horizontal gauge bosons have to be very heavy so that the
experimental constraints on the flavour changing neutral processes are
satisfied. In a general horizontal symmetric model CP-violation can occur via
complex phases in the charged weak currents as well as those in the horizontal
gauge boson couplings. If the model contains both these sources of CP-violation
it is impossible to do a meaningful analysis of the CP-violating observables
because too many undetermined parameters enter the calculation. It is then
possible to fine-tune these parameters in such a way that the predictions for
the CP-violating observables are commensurate with the experimental values and
the model has no predictive power. It is desirable to limit the CP-violation to
the horizontal sector alone and see what such a scenario would predict.

The four-fermion coupling due to horizontal interactions, $G_H$, is fixed
mainly by the requirement that the horizontal gauge boson contribution to
$\Delta m_K$ should not exceed the experimental value of $3.5 \times 10^{-15} \
GeV$.  It is interesting to note that this constraint requires $G_H$ to be of
superweak strength ($G_H \sim 10^{-8} G_F$) \cite{lw64}.  Decker, Gerard and
Zoupanos have calculated \cite{dgz} the CP-violating observables in general
horizontal symmetric models.  Their analysis suggests that, in models in which
the CP-violation is confined horizontal sector alone, it is possible to obtain
the correct value of $\eps$ while $|\epsp / \eps| \leq 10^{-3}$ and $d_n \geq
10^{-28}$ e-cm.

Several horizontal symmetries have been proposed to understand the problem of
fermion generations, fermion masses and the Kobayashi-Maskawa (KM) matrix. It
was shown that $SU(2)_H$ can lead to calculable quark mixing angles and
CP-violating phases \cite{fwaz} but $SU(3)_H$ fails \cite{drtj}.  Bandyopadhyay
and Ray have considered the relation between various horizontal symmetries and
fermion mass matrices \cite{kbar}.  They have demonstrated that Fritzsch type
mass matrices \cite{hf} and calculable quark mixing angles can be obtained in
models with the horizontal symmetry group $SU(3)_H^{VL}$ ($VL$ stands for
vector-like).  Under $SU(3)_H^{VL}$ the left-handed fermions transform as
triplets and the right-handed fermions transform as antitriplets. They have
explicitly constructed an attractive model based on the gauge group $SU(2)_L
\times SU(2)_R \times U(1)_{B-L} \times SU(3)_H^{VL}$.  In this model not only
are the quark mixing angles calculable but the KM matrices in both the left and
right handed sectors are real. Hence no CP-violation can occur due to charged
current interactions mediated by $W_L$ or $W_R$, and all CP-violating
interactions take place due to horizontal gauge boson exchange only. In a
previous paper \cite{brr} the $\eps$ parameter of the $K^0 - \bar{K}^0$ system
was calculated in this model in an approximation which simplifies the algebra
immensely.  For this case, it was found that the experimental value of $\eps$
can be fitted by fine-tuning the phases in the horizontal gauge boson mass
matrix.

In this paper we investigate the predictions of the general $SU(2)_L \times
SU(2)_R \times U(1)_{B-L} \times SU(3)_H^{VL}$ model for the CP-violating
observables $\eps$, $\epsp / \eps$ and $d_n$.  In section 2, we give a brief
description of the model. In section 3, we discuss CP-violation in the $K^0 -
\bar{K}^0$ system and present our results on $\eps$ and $\epsp$. Section 4 is
devoted to the neutron electric dipole moment $d_n$. Our conclusions are
presented in section 5 and some important expression are given in the Appendix.

\begin{center}  {\bf 2. The Model}  \end {center}

The model we are analyzing here is a horizontal symmetric model based on the
gauge group $SU(2)_L \times SU(2)_R \times U(1)_{B-L} \times SU(3)_H^{VL}$. Its
important features are
\begin{enumerate}
\item
unlike the usual left-right symmetric models, flavour changing
higgs couplings are absent at tree level;
\item
the fermion mass matrices are of Fritzsch type, leading to calculable
fermion masses and mixing angles;
\item
CP-violation is confined to the horizontal sector alone.
\end{enumerate}
The model is described in detail in Ref. \cite{kbar}.
Here we describe some of the essential features of the model. Under the
gauge group the left-handed fermions transform as $\left( 2, 1, \frac{1}{3},
3 \right)$ and the right-handed fermions transform as $\left( 1, 2,
\frac{1}{3}, \bar{3} \right)$. As in the left-right symmetric model there
are three $SU(2)_L$ gauge bosons $W^a_{\mu L}$, three $SU(2)_R$ gauge bosons
$W^a_{\mu R}$ and the $U(1)_{B-L}$ gauge boson $B_{\mu}$. In addition,
there are eight horizontal gauge bosons $G^a_{\mu} (a = 1,2,...,8)$ forming
an octect of $SU(3)_H^{VL}$. The minimal set of scalar multiplets required
to break horizontal, left-right and electroweak symmetries spontaneously, is
given by $\rho (1, 1, 0, 3)$, $\Delta_L (3, 1, 2, 1)$,
$\Delta_R (1, 3, 2, 1)$ and $\omega (2, 2, 0, 6)$, where the numbers in the
brackets denote the quantum numbers of various multiplets under the groups
$SU(2)_L, SU(2)_R, U(1)_{B-L}$ and $SU(3)_H^{VL}$ respectively. The symmetry
breaking in this model takes place according to the following scheme:
\begin{enumerate}
\item
The vacuum expectation values (VEVs) of the components of $\rho$ break the
horizontal symmetry and give rise to the masses of horizontal gauge bosons;
\item
the VEVs of the neutral components of $\Delta_L$ and $\Delta_R$ break the
left-right and the electroweak symmetries in such a way as to make
$W_{R \mu}$ much more massive than $W_{L \mu}$;
\item
the VEVs of the neutral components of $\omega$ generate the fermion
masses and also contribute to the masses of $W_{L \mu}$ and $W_{R \mu}$.
\end{enumerate}
In order to obtain the correct masses for the gauge bosons we must have
the following values for the VEVs of the various scalar multiplets,
\bea
\langle \rho \rangle & \sim & 10 \ TeV,  \nn   \\
\langle \Delta_R \rangle & \sim & 1 \ TeV, \nn \\
\langle \Delta_L \rangle & \sim & 100 \ GeV.
\eea
The VEVs of $\omega$ give rise to the masses
of various fermions and are of the general form
\ben
\langle \omega_{ij} \rangle = \left(
\begin{array}{cc}
k_{ij} e^{i \theta_{ij}} & 0 \\ 0 & k'_{ij} e^{i \theta'_{ij}} \\
\end{array} \right),
\label{eq:omvev}
\een
where $i$ and $j$ are generation indices.
In order to obtain calculable quark mixing angles, the VEVs in
equation~(\ref{eq:omvev}) are restricted such that the charge
$\frac{2}{3}$ quark
mass matrix ${\bf M}_u$ and the charge $-\frac{1}{3}$ quark mass matrix
${\bf M}_d$ are of the Fritzsch type \cite{hf}.
In Ref. \cite{kbar}
it has been demonstrated that the minimization of the most general
Higgs potential requires that
\ben
\theta_{ij} - \theta'_{ij} = 0 \ {\rm or} \ \pi.
\een
The complex symmetric matrices ${\bf M}_u$ and ${\bf M}_d$ can be
diagonalized by biunitary transformations consisting of unitary matrices
$U_u^L$, $U_u^R$ and $U_d^L$, $U_d^R$
respectively. Becasuse of the symmetry of ${\bf M}_u$ and ${\bf M}_d$
we have
\bea
U_u^L & = & U_u^{R^*} = {\bf F}_u \ {\bf O}_u,
\label{eq:fuou}  \\
U_d^L & = & U_d^{R^*} = {\bf F}_d \ {\bf O}_d.
\label{eq:fdod}
\eea
In equations~(\ref{eq:fuou}) and~(\ref{eq:fdod}), the matrices ${\bf O}_u$ and
${\bf O}_d$ are $3 \times 3$ orthogonal matrices (see Appendix)
and ${\bf F}_u$ and ${\bf F}_d$ are diagonal phase matrices of
the form
\ben
{\bf F}_{u,d} = \left( \begin{array}{ccc}
\exp \left( i \xi_{u,d} \right) & 0 & 0 \\
0 & \exp \left( i \xi_{c,s} \right) & 0 \\
0 & 0 & \exp \left( i \xi_{t,b} \right)  \\
\end{array} \right).
\een
Since the phases appearing in the mass matrices ${\bf M}_u$ and
${\bf M}_d$ are equal to each other (or differ by $\pi$), the
corresponding phases in the matrices ${\bf F}_u$ and ${\bf F}_d$
are also equal to each other (or differ by $\pi$). Therefore the
KM matrices in the left and the right handed sectors
turn out to be real and identical, since
\bea
V_L & = & U_u^{L^{\dagger}} U_d^L = {\bf O}_u^T {\bf O}_d
\nn    \\
V_R & = & U_u^{R^{\dagger}} U_d^R = {\bf O}_u^T {\bf O}_d.
\eea
Hence there is no CP-violation in the charged current sector in this model.

In the horizontal sector, we have the following mass matrix for the eight
horizontal gauge bosons after the spontaneous symmetry breaking,
\ben
{\bf M}_{ab} = \frac{g_H^2}{4} \sum_{i=1}^3 \sum_{j=1}^3 v_i v_j \exp \left[ i
\left( \beta_i - \beta_j \right) \right] \left( \lambda_a \lambda_b
\right)_{ij},
\label{eq:hgbmm}
\een
where $a,b$ are $SU(3)_H$ indices and run from $a,b = 1,2,...,8$ and
$\lambda_a$ are the Gell-Mann matrices. In equation~(\ref{eq:hgbmm}), $g_H$ is
the coupling constant for the horizontal boson interactions and $v_i$ and
$\beta_i$ are respectively the magnitudes and the phases of the VEVs of the
components of $\rho$,
\ben
\langle \rho_i \rangle = v_i \exp \left( i \beta_i \right).
\een
The expanded form of ${\bf M}_{ab}$ is given in the Appendix.
The absolute values of
the VEVs are related to the four-fermion coupling $G_H$ by
\bea
v_1^2 + v_2^2 + v_3^2 = v^2 & = & \left( \sqrt{2} G_H \right)^{-1} \nn \\ & = &
\frac{4 M_H^2}{g_H^2}. \label{eq:defmh}
\eea
$M_H$ in equation~(\ref{eq:defmh}), is the scale of the horizontal
gauge boson mass which we take to be $10 \ TeV$.  Physical horizontal gauge
bosons, which we denote by $H_{\mu}^a$, correspond to the eigenstates of the
mass matrix ${\bf M}_{ab}$ and are given by
\ben
H_{\mu}^a = \sum_{b=1}^8 {\bf C}^{ab} G_{\mu}^b,
\een
where the unitary matrix ${\bf C}$ diagonalizes the hermitian matrix ${\bf M}$.

The interactions between the horizontal bosons and the charge
$-\frac{1}{3}$ quarks are
given by
\ben
{\cal L}_H = g_H \sum_{a=1}^8 \left[ \bar{D^0}_L \gamma^{\mu}
\frac{\lambda^a}{2} D^0_L G_{\mu}^a - \ ( L \ra R ) \right] + H.c.,
\label{eq:hziges}
\een
where $D^0_{L,R} = \left( d^0, s^0, b^0 \right)_{L,R}^T$ is the weak
eigenbasis.  On rewriting the interaction term in equation~(\ref{eq:hziges}) in
terms of the quark mass eigenbasis $D = (d, s, b)^T$ and the physical
horizontal bosons $H^a$, we obtain
\ben
{\cal L}_H = g_H \sum_{a=1}^8 \bar{D} \gamma_{\mu} \left( {\bf A}^a + {\bf B}^a
\gamma_5 \right) D H^a_{\mu} + H.c.
\label{eq:hzimes}
\een
The matrices ${\bf A}^a$ and ${\bf B}^a$ are given by
\bea
{\bf A}^a  & = & - i {\bf O}_d^{\dagger} {\bf S}^a {\bf O}_d  \nn  \\
{\bf B}^a  & = & - {\bf O}_d^{\dagger} {\bf T}^a {\bf O}_d,
\label{eq:defAB}
\eea
where
\bea
{\bf S}^a_{ij} & = & \sum_{b=1}^8 \left( {\bf C}^{\dagger} \right)^{ab} \left(
\lambda^b \right)_{ij} \sin \left( \xi_i - \xi_j \right), \nn  \\
{\bf T}^a_{ij} & = & \sum_{b=1}^8 \left( {\bf C}^{\dagger} \right)^{ab} \left(
\lambda^b \right)_{ij} \cos \left( \xi_i - \xi_j \right).
\label{eq:defCS}
\eea
The matrices ${\bf A}$ and ${\bf B}$ in equation~(\ref{eq:hzimes}), which
define the quark-horizontal gauge boson interactions in the physical
eigenbasis, contain six phases $\xi_i$ and $\beta_i$.  Of these
$\xi_3$ and $\beta_3$ can be absorbed in the overall phases of the quark and
the horizonatal gauge boson fields respectively.  The other phases can have
non-vanishing values and cause CP-violation in
interactions mediated by the horizontal gauge bosons.

{}From the complete form of mass matrix ${\bf M}_{ab}$ given in the Appendix we
observe that if the VEV $v_3$ is chosen to be zero, ${\bf M}_{ab}$ reduces to a
block diagonal form consisting of two $4 \times 4$ blocks, one consisting of
$( G_1, G_2, G_3, G_8 )$ and the other consisting of $( G_4, G_5, G_6, G_7 )$.
This simplification was made in Ref. \cite{brr} in the calculation of $\eps$.
Here, on the other hand, we have considered the whole $8 \times 8$ form for
${\bf
M}_{ab}$ with $v_3 \neq 0$ and diagonalized the matrix numerically for various
choices of $v_1, v_2, v_3$. We find that the approximation $v_3 = 0$, while
simplifying the algebra, also drastically reduces the allowed parameter
space satisfying the constraint on $\eps$.

\begin{center}
{\bf 3. CP-violation in $K^0 - \bar{K}^0$ System}
\end{center}

The off-diagonal terms in the matrices ${\bf A}$ and ${\bf B}$
in equation~(\ref{eq:hzimes}) lead to flavour changing neutral
interactions at tree level. Hence we have  $K^0 \leftrightarrow
\bar{K}^0$ transition via the diagram shown in Fig. 1(a).
This diagram gives rise to an effective $\Delta S = 2$ hamiltonian given by
\ben
{\cal H}_{eff}^{\Delta S = 2} = g_H^2 \sum_{a=1}^8 \frac{1}{M_a^2}
\bar{s} \gamma^{\mu} \left( {\bf A}^a_{21} + {\bf B}^a_{21} \gamma_5 \right) d~
\bar{s} \gamma_{\mu} \left( {\bf A}^{a^*}_{12} + {\bf B}^{a^*}_{12}
\gamma_5 \right) d.
\label{eq:heffds2}
\een
Evaluating the matrix element of this hamiltonian
between $K^0$ and $\bar{K}^0$ states in the vacuum saturation approximation we
obtain
\bea
{\cal M}_{12}^H & = & \frac{4 G_H f_K^2 m_K B_K}{3 \sqrt{2}} \sum_{a=1}^8
\frac{M_H^2}{M_a^2} \left[ \left( {\bf A}^a_{21} {\bf A}^{a^*}_{12} + 7 {\bf
B}^a_{21} {\bf B}^{a^*}_{12} \right) \right.
\nn \\
      & & \left. - 2 \left( \frac{m_K}{m_d + m_s} \right)^2 \left( {\bf
A}^a_{21} {\bf A}^{a^*}_{12} - {\bf B}^a_{21} {\bf B}^{a^*}_{12} \right)
\right].
\label{eq:dmkhz}
\eea
The real part of ${\cal M}_{12}^H$ contributes to the $K_L - K_S$ mass
difference whereas the imaginary part of it gives rise to $\eps$,
\bea
\Delta m_K^H & = & 2 \ Re \left( {\cal M}_{12}^H \right),   \\  \nn
\epsilon & = & \left( \frac{ Im \  {\cal M}_{12}^H}{\sqrt{2} \Delta m_K}
\right).
\eea
The experimental value of the $K_L - K_S$ mass difference is $\Delta m_K = 3.5
\times 10^{-15} \ GeV$.  We have used the experimental value of $\Delta m_K$ in
the expression for $\eps$ because all theoretical calculations of $\Delta m_K$
are subject to large uncertainties.

In this model $\epsp$ also occurs at tree level because the horizontal gauge
bosons can mediate $\Delta S = 1$ neutral current interactions.  The relevant
diagram is shown in Fig. 1(b). The effective hamiltonian is given by
\ben
{\cal H}_{eff}^{\Delta S = 1} = g_H^2 \sum_{a=1}^8 \frac{1}{M_a^2}
\bar{s} \gamma^{\mu} \left( {\bf A}^a_{21} + {\bf B}^a_{21} \gamma_5 \right) d~
\bar{d} \gamma_{\mu} \left( {\bf A}^{a^*}_{11} + {\bf B}^{a^*}_{11}
\gamma_5 \right) d.
\label{eq:heffds1}
\een
The matrix element of this hamiltonian between kaon and pion states is
\bea
{\cal M}_{K^0 \ra \pi^0}^H & = & \frac{G_H f_{\pi} f_K m_K}{3} \sum_{a=1}^8
\frac{M_H^2}{M_a^2} \left[ \left( {\bf A}^a_{21} {\bf A}^{a^*}_{11} + 7 {\bf
B}^a_{21} {\bf B}^{a^*}_{11} \right) \right.
\nn \\
      & & \left. - \left( \frac{m_{\pi}^2}{m_d (m_d + m_s)} \right) \left( {\bf
A}^a_{21} {\bf A}^{a^*}_{11} - {\bf B}^a_{21} {\bf B}^{a^*}_{11} \right)
\right].
\label{eq:ak2pi}
\eea
Relating the $K^0 \ra \pi^0$ amplitude in equation~(\ref{eq:ak2pi})
to the $K^0 \ra 2 \pi$ amplitude via the soft pion theorem,
\ben
{\cal M}^H_{K^0 \ra 2 \pi} = - \frac{i}{\sqrt{2} f_{\pi}} {\cal M}^H_{K^0 \ra
\pi^0},
\een
we estimate $\epsp$ as \cite{fgjh}
\ben
| \epsp | = \frac{1}{\sqrt{2}} \left| \frac{A_2}{A_0} \right|
\frac{ | Im \ {\cal M}^H_{K^0 \ra 2 \pi} | }{(Re \
           {\cal M}_{K^0 \ra 2 \pi})_{exp}},
\label{eq:epspest}
\een
where $|A_2/A_0| \sim 1/22$ is the ratio of $\Delta I = 3/2$ to
$\Delta I = 1/2$ amplitudes in the $K^0 \ra 2 \pi$ transition.
{}From the $K_S \ra 2 \pi$ decay rate, we obtain
\ben
(Re \ {\cal M}_{K^0 \ra 2 \pi})_{exp} \sim 3 \times 10^{-7} \ GeV.
\een
We used the experimental value of $Re \ {\cal M}_{K^0
\ra 2 \pi}$ to estimate $\epsp$ because
there is no reliable theoretical method
to calculate $K^0 \ra 2 \pi$ decay widths.
Finally we estimate $| \epsp / \eps |$ by dividing the value obtained from
equation~(\ref{eq:epspest}) by the experimental value of $\eps$.

We have calculated the effect of the horizontal gauge boson interactions on
$B^0_d - \bar{B}^0_d$ mixing. The contribution to $\Delta m_B$ from the
horizontal boson exchange diagram for most choices of parameters is two to
three orders of magnitude smaller than the experimental value.

\begin{center}     {\bf 4. Electric Dipole Moment of Neutron}   \end{center}

The electric dipole moment of the neutron is taken to be the sum of the
electric dipole moments of its constituent quarks and is given by
\ben
d_n = \frac{4}{3} \mu (d) - \frac{1}{3} \mu (u).
\een
The electric dipole moment of the $d$-quark is obtained from the diagram shown
in Fig. 1(c). Evaluating this diagram we get
\ben
\mu (d) = \frac{\sqrt{2} \ e \ G_H}{3 \pi^2} \sum_{a=1}^{8} \sum_{i=1}^3
          m_i \left( \frac{M_H}{M_a} \right)^2 f \left( \frac{m_i^2}{M_a^2}
\right) Im \ \left( {\bf A}_{1i} {\bf B}_{i1}^* \right),
\label{eq:demon}
\een
where $m_i$ is the mass of the $i$th $-\frac{1}{3}$ charge quark.
The argument of the function
$f$ in equation~(\ref{eq:demon}) is very small $(m_i^2 \ll M_a^2)$. In this
limit we can use the approximation
\ben
f (x) \simeq \ln x.
\een
We have not calculated $\mu (u)$ as this requires further analysis of the
coupling of the charge $\frac{2}{3}$ quarks to the horizontal gauge bosons.
This leads
to the introduction of many more parameters of the model into the calculation.
This does not make the discussion any more illuminating because it is expected
that $\mu (u)$ and $\mu (d)$ are similar in magnitude and hence $d_n$ is of the
same order of magnitude as $\mu (d)$.

\begin{center}  {\bf 5. Discussion of Results and Conclusion}   \end{center}

      From the formulae in the text, we see that all the CP-violating
observables scale as $G_H$, but different choices of the VEVs $v_1, v_2, v_3$
could potentially lead to widely different values of these.
This variation cannot be
studied analytically because it arises from the diagonalization of the $8
\times 8$ mass matrix for the horizontal gauge bosons. We have therefore
diagonalized the matrix numerically for various choices of $v_1, v_2, v_3$ and
found that, in fact, $\eps$ varies over two orders of magnitude. To obtain a
prediction for $\eps$ commensurate with the experimental value, we have chosen
an appropriate $G_H$ for each choice of the VEVs, and then calculated the
CP-violating observables for $0 \leq \beta_1, \beta_2
\leq 360^0$. The variation of $\eps$ with $\beta_1$ and $\beta_2$
is plotted in Fig. 2. Since our selection of $G_H$ was somewhat arbitrary, we
considered all values of $\beta_1, \beta_2$ which yielded the correct order of
magnitude for $\eps$.  From the graph one can see that the variation in $\eps$
is within an order of magnitude for the whole $(\beta_1, \beta_2)$ plane,
except for the lines characterized by $(\beta_1 - \beta_2) =
(2n+1) 90^0$.  This justifies our procedure of choosing an order of magnitude
value of $G_H$ for each set of $v_1, v_2, v_3$.

    Some representative values of $v_1, v_2, v_3$ chosen by us and the
corresponding values of $G_H$ are listed in Table 1, together with our
predictions for horizontal gauge boson contributions to $\Delta m_K ,\Delta m_B
,\eps, \epsp$ and $d_n$. It is obvious from a glance at the Table that $\Delta
m_K^H, \Delta m_B^H$ are smaller than the experimental values.
In fact, we notice that $\Delta m_K^H / \Delta m_K^{exp}$ is
similar in magnitude to $\eps$. This is not surprising because
the denominators of the two ratios are the same and the numerators
are respectively the real and imaginary parts of the same matrix
element (equation~(\ref{eq:dmkhz})).  The observed $K^0 - \bar{K}^0$
and $B^0_d - \bar{B}_d^0$ mixings may be attributed to the usual
box diagrams with $W_L$ (or $W_R$) exchange. When $G_H$ is
adjusted to give $\eps$ of the right order $\epsp / \eps$ comes out to be five
to six orders of magnitude below the current experimental upper bound for the
entire parameter space. This, of course, is just a manifestation of the
superweak nature of the model.  For the entire parameter space, $d_n$ comes out
to be $10^{-31} e-cm$ which is comparable to the SM prediction.

    All the CP-violating observables depend on the values of the nonzero
phases in the model, viz. $\xi_1, \xi_2, \beta_1, \beta_2$.  From equations
(\ref{eq:defAB}) and (\ref{eq:defCS}) the dependence of the flavour-changing
couplings on $\xi_1$ and $\xi_2$ is manifest. The CP-violating observables are
found to vary somewhat weakly with $\xi_1$ and $\xi_2$. We, therefore, fixed
values $\xi_1 = 30^0$ and $\xi_2 = 225^0$ in all our calculations. Other
choices of $\xi_1, \xi_2$ do not affect our conclusions significantly. The
dependence on $\beta_1$ and $\beta_2$ is hidden in the numerical complexities
of our approach. Hence we calculated $\eps$, $\epsp$ and $d_n$ for all values
of
$\beta_1$ and $\beta_2$ and found that their variation is limited to
within an order of
magnitude in the entire $(\beta_1, \beta_2)$ plane.

       To conclude we have analyzed the CP-violating observables in the
$SU(2)_L \times SU(2)_R \times \times U(1)_{B-L} \times SU(3)_H^{VL}$ model for
all choices of the parameters. When $G_H$, the four-fermion coupling
corresponding to horizontal interactions, is fixed to give the correct value of
$\eps$ it is found that $\epsp / \eps$ is strongly suppressed $(\leq 10^{-8})$
irrespective of the choice of phases. The neutron electric dipole moment $d_n$
also turns out to be very small (comparable to the SM prediction of $10^{-31}
e-cm$). It is apparent, therefore that though this model provides an attractive
scenario for CP-violation, the CP-violating observables do not provide a test
that can distinguish it from the other superweak models. The only conclusion
that can be drawn is that if experiments discover nonzero values of either
$\epsp / \eps$ or $d_n$ just below the current upper bounds, then this model
can be conclusively ruled out.

\underline{ACKNOWLEDGEMENTS:} We thank K. Bandyopadhyay,
D. Chatterjee, A. Datta and A. Raychaudhuri for helpful discussions.  The
authors are grateful to the Department of Science and Technology, India
(project no. SP/SP2/K32/89) for financial support. D. B. and S. R. would linke
to thank University Grants Commision, India for funding.

\newpage
\begin{center}   {\bf Appendix}   \end{center}

The orthogonal matrix which diagonalizes the mass matrix of the charge
$-\frac{1}{3}$ quarks is
\ben
{\cal O}_d = \left( \begin{array}{ccc} 1 & - \left( \frac{m_d}{m_s} \right) & 0
\\
\left( \frac{m_d}{m_s} \right) & 1 & \left( \frac{m_s}{m_b} \right) \\
- \left( \frac{m_d}{m_b} \right) & - \left( \frac{m_s}{m_b} \right) & 1
\end{array} \right).
\label{eq:formod}
\een

The mass matrix for the horizontal gauge bosons is obtained through the VEVs of
the Higgs multiplet $\rho$. These VEVs are assumed to be of the form
\ben
\langle \rho_i \rangle = v_i exp ( i \beta_i ).
\een
The relation between $G_H$ and the $v_i$'s is given in
equation~(\ref{eq:defmh}). For convenience we define the following quantities
\bea
r_i & = & \frac{v_i}{\sqrt{v_1^2 + v_2^2 + v_3^2}} \nn \\
\beta_{ij} & = & \beta_i - \beta_j. \nn
\eea
The complete $8 \times 8$ mass matrix for the horizontal gauge bosons is
\ben
{\cal M} = \frac{1}{4} g_H^2 \left( v_1^2 + v_2^2 + v_3^2 \right)
\left( \begin{array}{cc} {\cal M}_A & {\cal M}_B \\
{\cal M}_B^{\dag} & {\cal M}_C \end{array} \right),
\een
where
\ben
{\cal M}_A = \left( \begin{array}{cccc}
\left( r_1^2 + r_2^2 \right) &
i \left( r_1^2 - r_2^2 \right) & 2 i r_1 r_2 \sin \beta_{12} & r_2 r_3 exp ( -
i \beta_{23} ) \\ - i \left( r_1^2 - r_2^2 \right) &
\left( r_1^2 + r_2^2 \right) &
2 i r_1 r_2 \cos \beta_{12} & i r_2 r_3 exp ( - i \beta_{23} ) \\ - 2 i r_1 r_2
\sin \beta_{12} & - 2 i r_1 r_2 \cos \beta_{12} &
\left( r_1^2 + r_2^2 \right) &
r_1 r_3 exp ( i \beta_{31} ) \\ r_2 r_3 exp ( i\beta_{23} ) & - i r_2 r_3 exp (
i \beta_{23} ) & r_1 r_3 exp ( - i \beta_{31} ) &
\left( r_1^2 + r_3^2 \right) \\
\end{array} \right),
\label{eq:MA}
\een
\ben
{\cal M}_B = \left( \begin{array}{cccc} - i r_2 r_3 exp ( - i \beta_{23} ) &
r_1 r_3 exp ( i \beta_{31} ) & - i r_1 r_3 exp ( i \beta_{31} ) &
\frac{2}{\sqrt{3}} r_1 r_2 \cos \beta_{12} \\
r_2 r_3 exp ( - i \beta_{23} )  & i r_1 r_3 exp ( i \beta_{31} ) & - r_2 r_3
exp ( i \beta_{31} ) & - \frac{2}{\sqrt{3}} r_1 r_2 \sin \beta_{12} \\ - i r_1
r_3 exp ( i \beta_{31} ) & - r_2 r_3 exp ( - i \beta_{23} ) & i r_2 r_3 exp ( -
i \beta_{23} ) &
\frac{1}{\sqrt{3}} \left( r_1^2 - r_2^2 \right) \\
i \left( r_1^2 - r_3^2 \right) & r_1 r_2 exp (- i \beta_{12} ) & i r_1 r_2 exp
( - i \beta_{12} ) &
\frac{1}{\sqrt{3}} r_1 r_3 [ exp (i \beta_{31}) \\
 & & & - 2 exp ( - i \beta_{31} ) ]
\end{array} \right),
\label{eq:MB}
\een
\ben
{\cal M}_C = \left( \begin{array}{cccc}
\left( r_1^2 + r_3^2 \right) &
-i r_1 r_2 exp ( - i \beta_{12} ) & r_1 r_2 exp ( - i \beta_{12} ) & i r_1 r_3
[ 2 exp ( i \beta_{31}) \\ & & & + exp ( - i \beta_{31} ) ]    \\ i r_1 r_2 exp
( i \beta_{12} ) &
\left( r_2^2 + r_3^2 \right) &
i \left( r_2^2 - r_3^2 \right) &
\frac{1}{\sqrt{3}} r_2 r_3 [ exp ( i \beta_{23}) \\
 & & & - 2 exp ( - i \beta_{23} ) ]     \\ r_1 r_2 exp ( i \beta_{12} ) & - i
\left( r_2^2 - r_3^2 \right) &
\left( r_2^2 + r_3^2 \right) &
\frac{i}{\sqrt{3}} r_2 r_3 [ exp ( i \beta_{23}) \\
 & & & - 2 exp ( - i \beta_{23} ) ]     \\ - i r_1 r_3 [ 2 exp ( - i
\beta_{31}) &
\frac{1}{\sqrt{3}} r_2 r_3 [ exp ( - i \beta_{23}) &
- \frac{i}{\sqrt{3}} r_2 r_3 [ exp ( - i \beta_{23}) &
\frac{1}{3} \left( r_1^2 + r_2^2 \right. \\
+ exp ( i \beta_{31} ) ] & - 2 exp ( i \beta_{23} ) ]  & - 2 exp ( i \beta_{23}
) ]  &
\left. + 4 r_3^2 \right)
\end{array} \right).
\een

\newpage
\begin{table}
\begin{tabular}{|c|c|c|c|c|c|c|c|}
\hline
$r_1$ & $r_2$ & $r_3$ & $G_H$ (in GeV$^{-2}$) & $\frac{\Delta m_K^H}{\Delta
m_K^{exp}}$ & $\frac{\Delta m_B^H}{\Delta m_B^{exp}}$ & $|\frac{\epsp}{\eps}|$
& $d_n$ (in e-cm)\\
\hline
$\frac{1}{\sqrt{3}}$ & $\frac{1}{\sqrt{3}}$ & $\frac{1}{\sqrt{3}}$ & $5 \times
10^{-16}$ & $10^{-2}$ & $10^{-3}$ & $10^{-9}$ & $10^{-31}$  \\
\hline
$\sqrt{0.2}$ & $\sqrt{0.3}$ & $\sqrt{0.5}$ & $5 \times 10^{-16}$ &
$10^{-2}$ & $10^{-3}$ & $10^{-9}$ & $10^{-31}$ \\
\hline
$0.01$ & $0.1$ & $0.995$ & $5 \times 10^{-14}$ & $10^{-1}$ & $10^{-1}$ &
$10^{-8}$ & $10^{-30}$ \\
\hline
$0.05$ & $0.707$ & $0.707$ & $5 \times 10^{15}$ & $10^{-2}$ & $10^{-2}$ &
$10^{-8}$ & $10^{-30}$ \\
\hline
$0.01$ & $0.99$ & $0.141$ & $5 \times 10^{-14}$ & $10^{-1}$ & $10^{-1}$ &
$10^{-8}$ & $10^{-30}$ \\
\hline
$0.1$ & $0.99$ & $0.1$ & $10^{-15}$ & $10^{-3}$ & $10^{-2}$ & $10^{-9}$ &
$10^{-31}$ \\
\hline
$0.141$ & $0.99$ & $0.01$ & $10^{-15}$ & $10^{-3}$ & $10^{-3}$ & $10^{-9}$ &
$10^{-31}$ \\
\hline
$0.12$ & $0.15$ & $0.98$ & $10^{-15}$ & $10^{-3}$ & $10^{-3}$ & $10^{-9}$ &
$10^{-31}$ \\
\hline
$0.799$ & $0.599$ & $0.05$ & $10^{-15}$ & $10^{-2}$ & $10^{-2}$ & $10^{-9}$ &
$10^{-30}$ \\
\hline
$0.707$ & $0.707$ & $0.004$ & $5 \times 10^{-15}$ & $10^{-3}$ & $10^{-3}$ &
$10^{-9}$ & $10^{-31}$ \\
\hline
$0.99$ & $0.01$ & $0.141$ & $10^{-15}$ & $10^{-3}$ & $10^{-3}$ & $10^{-8}$ &
$10^{-30}$ \\
\hline
$0.7$ & $0.01$ & $0.714$ & $5 \times 10^{-16}$ & $10^{-2}$ & $10^{-2}$ &
$10^{-9}$ & $10^{-31}$ \\
\hline
$0.1$ & $0.01$ & $0.995$ & $5 \times 10^{-16}$ & $10^{-3}$ & $10^{-3}$ &
$10^{-9}$ & $10^{-31}$ \\
\hline
$0.99$ & $0.1$ & $0.1$ & $10^{-15}$ & $10^{-3}$ & $10^{-2}$ & $10^{-9}$ &
$10^{-30}$ \\
\hline
$0.99$ & $0.141$ & $0.01$ & $10^{-15}$ & $10^{-3}$ & $10^{-3}$ & $10^{-10}$ &
$10^{-30}$ \\
\hline
\end{tabular}
\caption{Values of $G_H$, $\Delta m_K^H$, $\Delta m_B^H$, $\epsp$ and $d_n$
for various choices of $v_1$, $v_2$ and $v_3$.}
\end{table}
\phantom{Pakistan is No.1}

\newpage

\begin{center}
\begin{picture}(200,150)
\put(0,50){\line(1,1){50}}
\put(25,75){\vector(1,1){0}}
\put(33,75){$s$}
\put(50,100){\line(-1,1){50}}
\put(25,125){\vector(-1,1){0}}
\put(33,125){$d$}
\put(200,150){\line(-1,-1){50}}
\put(175,125){\vector(-1,-1){0}}
\put(167,125){$s$}
\put(150,100){\line(1,-1){50}}
\put(175,75){\vector(1,-1){0}}
\put(160,75){$d$}
\put(60,100){\oval(20,20)[t]}
\put(80,100){\oval(20,20)[b]}
\put(100,100){\oval(20,20)[t]}
\put(120,100){\oval(20,20)[b]}
\put(140,100){\oval(20,20)[t]}
\put(100,80){$H^a$}
\put(100,0){\makebox(0,0){Fig. 1(a)}}
\end{picture}
\end{center}

\vspace{0.5cm}

\begin{center}
\begin{picture}(200,150)
\put(0,50){\line(1,1){50}}
\put(25,75){\vector(1,1){0}}
\put(33,75){$s$}
\put(50,100){\line(-1,1){50}}
\put(25,125){\vector(-1,1){0}}
\put(33,125){$d$}
\put(200,150){\line(-1,-1){50}}
\put(175,125){\vector(-1,-1){0}}
\put(167,125){$d$}
\put(150,100){\line(1,-1){50}}
\put(175,75){\vector(1,-1){0}}
\put(160,75){$d$}
\put(60,100){\oval(20,20)[t]}
\put(80,100){\oval(20,20)[b]}
\put(100,100){\oval(20,20)[t]}
\put(120,100){\oval(20,20)[b]}
\put(140,100){\oval(20,20)[t]}
\put(100,80){$H^a$}
\put(100,0){\makebox(0,0){Fig. 1(b)}}
\end{picture}
\end{center}

\vspace{1cm}

\begin{center}
\begin{picture}(240,200)
\put(0,180){\line(1,0){70}}
\put(35,180){\vector(1,0){0}}
\put(30,190){$d$}
\put(170,180){\line(1,0){70}}
\put(205,180){\vector(1,0){0}}
\put(200,190){$d$}
\put(80,180){\oval(20,20)[t]}
\put(100,180){\oval(20,20)[b]}
\put(120,180){\oval(20,20)[t]}
\put(140,180){\oval(20,20)[b]}
\put(160,180){\oval(20,20)[t]}
\put(130,190){$H^a$}
\put(70,180){\line(1,-1){50}}
\put(95,155){\vector(1,-1){0}}
\put(120,130){\line(1,1){50}}
\put(145,155){\vector(1,1){0}}
\put(120,145){$d_j$}
\put(120,120){\oval(20,20)[l]}
\put(120,100){\oval(20,20)[r]}
\put(120,80){\oval(20,20)[l]}
\put(120,60){\oval(20,20)[r]}
\put(120,40){\oval(20,20)[l]}
\put(130,80){$\gamma$}
\put(120,0){\makebox(0,0){Fig. 1(c)}}
\end{picture}
\end{center}

\newpage

\begin{center} {\bf Figure Captions} \end{center}

\begin{enumerate}
\item
\begin{enumerate}
\item
Horizontal gauge boson exchange diagram for $K^0 - \bar{K}^0$ mixing.
\item
Horizontal gauge boson exchange diagram for $K \ra \pi$ transition.
\item
Feynman diagram leading to the electric dipole moment of the d-quark.
\end{enumerate}
\item
Variation of $\eps$ with $\beta_1$ and $\beta_2$. $\eps$ is plotted in units of
$10^{-3}$. $G_H$ is taken to be $5 \times 10^{-16} \ GeV^{-2}$ and $v_1 = v_2 =
v_3$.
\end{enumerate}

\vspace{1cm}

\begin{center}  {\bf Table Caption} \end{center}

\begin{enumerate}
\item
Values of $G_H$, $\Delta m_K^H$, $\Delta m_B^H$, $\epsp$ and $d_n$
for various choices of $v_1$, $v_2$ and $v_3$.
\end{enumerate}
\end{document}